\documentclass[twocolumn,aps,prl]{revtex4}% Physical Review Letter

\usepackage{graphicx}% Include figure files

\begin{document}

%\preprint{MgBC2}

\title{Carbon-substitution effect on the electronic properties of MgB$_2$ single crystals}
\author{T. Masui} 
%\altaffiliation{JSPS Research Fellow}
\email{masui@istec.or.jp}
\author{S. Lee}
\author{S. Tajima}

\affiliation{Superconductivity Research Laboratory, ISTEC, 1-10-13 Shinonome, Tokyo, 135-0062, Japan}
\date{\today}

\begin{abstract}

The electronic properties of the carbon substituted MgB$_2$ single crystals 
are reported. 
The carbon substitution drops T$_c$ below 2 K. 
In-plane resistivity shows a remarkable increase in residual resistivity 
by C-substitution, while 
the change of in-plane/out-of-plane Hall coefficients is rather small. 
Raman scattering spectra indicate that the E$_{2g}$-phonon frequency 
radically hardens with increasing the carbon-content, 
suggesting the weakening of electron-phonon coupling. 
Another striking C-effect is the increases of the second critical fields 
in both in-plane and out-of-plane directions, 
accompanied by a reduction in the anisotropy ratio. 
The possible changes in the electronic state and the origin of T$_c$-suppression by C-substitution are discussed. 
\end{abstract}


\maketitle

\section{Introduction}
The electronic properties of MgB$_2$  
have been intensively studied since the discovery  of superconductivity in this compound \cite{nagamatsu}. 
The band structure of MgB$_2$ is characterized by the two-dimensional $\sigma$-bands 
and the three-dimensional $\pi$-bands \cite{kortus,Uchiyama,Yelland,Eltsev_R_H}. 
The high superconducting transition temperature (T$_c$) originates from the 
strong electron-phonon (e-ph) coupling on the two-dimensional $\sigma$-bands \cite{an_pickett,Golubov,HJ_Choi}. 
A completely different strength of e-ph coupling in the $\sigma$-bands and 
$\pi$-bands results in two superconducting (SC) gaps. 
One is a larger gap on the $\sigma$-bands ($\Delta_\sigma$) 
and the other is a smaller one on the $\pi$-bands ($\Delta_\pi$) \cite{Bouquet_poly,Tsuda_poly,wang,quilty_c-axis,Bouquet_single,Souma,Tsuda_single}. 
The superconducting state shows a large anisotropy, 
reflecting  mainly the nature of  $\Delta_\sigma$ 
on cylindrical $\sigma$-bands 
\cite{Bouquet_single,Simon,Eltsev_Hc2,xu,kim,Pradhan_01,sologubenko}. 

Parallel with these studies for understanding a superconductivity mechanism, 
possibilities to control the electronic properties have been investigated, 
utilizing pressure, chemical substitution and irradiation. 
Among the studies, chemical substitution effect has been one of the hot topics 
\cite{Slusky_Al,Takenobu,Ribeiro}. 
Although there are many AlB$_2$-type compounds, 
systematic substitution-study from MgB$_2$ is limited to the study 
of Al-substitution for Mg 
and C-substitution for B \cite{Cava_Review}. 
The primary effect of Al-/C-substitution is a drop 
of T$_c$. 
We can list up possible sources for the T$_c$-suppression, 
such as a decrease in density of states, 
strong interband scattering\cite{Liu}, and the weakening of e-ph coupling. 
Although electron doping is expected from the valence consideration, 
the experimentally observed evidence is unclear. 
Furthermore, the observed decrease in anisotropy ratio of the second critical field (H$_{c2}$) indicates a substantial change of the electronic 
state \cite{Ribeiro,Pissas}. 
The C-substitution effect on the two-gap state is also interesting. 
The recent measurements on polycrystalline samples 
have revealed the existence of two gaps 
in the C-substituted MgB$_2$ with T$_c$ around 20 K \cite{Ribeiro,Samuely,Schmidt}, which is contrary to 
the theoretical prediction that the $\pi$- and $\sigma$-band gaps merge into a single gap giving $T_c$=25 K \cite{Liu}. 
This fact casts a doubt on the interband scattering scenario. 
In order to clarify these problems, 
it is necessary to study more systematically many physical properties 
using single crystals. 

In this paper we try to clarify the carbon-substitution effect on the 
electronic properties of MgB$_2$, using a series of C-substituted  
single crystals. 
Electrical resistivity and Raman scattering were measured 
to investigate the change of disorder effect and/or e-ph coupling.  
The residual part of in-plane resistivity dramatically increases 
with increasing carbon content, 
suggesting the increase of impurity scattering. 
On the other hand, 
the change in the slope $d\rho/dT$ is not strong with C-content, 
reflecting the complicated contribution from the multi-bands. 
In Raman scattering spectra, the E$_{2g}$-phonon peak shifts toward 
higher energy as the carbon-content increases, suggesting the weakening of 
e-ph coupling. 
The change in carrier density was examined by measuring 
Hall coefficient (R$_H$). 
The C-effect on in-plane/out-of-plane Hall coefficients is weak even 
in heavily C-substituted crystal, which agrees 
fairly well with the theoretical calculation within 
a rigid band scheme. 
In the superconducting state, anisotropy ratio decreases with increasing 
carbon content. 
The slope of H$_{c2}$-line ($dH_{c2}/dT$) increases with C-substitution, 
giving higher H$_{c2}$ in the moderately 
C-substituted crystals, in both magnetic field directions than 
that of C-free MgB$_2$. 
These changes are explained by the increase of impurity scattering and 
the modification of band structure by C-substitution. 

\section{Experiment}
Single crystals of MgB$_2$ were grown under high-pressure \cite{Lee_JPSJ}. 
The carbon-substituted MgB$_2$ were grown from mixture 
of Mg, amorphous B and C powders \cite{Lee_MGBC}. 
The carbon substitution up to 15 \% per boron atom was achieved.
Crystals were extracted mechanically from the bulk samples, 
and well characterized by a four-circle X-ray diffraction technique. 
The systematic change in the lattice parameter \cite{Lee_MGBC} and the 
small reliability factor R ($\leq$ 3 \%) obtained in the structure 
refinement \cite{Yamamoto_pr_com} prove that C-atoms are successfully substituted 
for B-atoms. 
%The C-contents were analyzed by an Auger spectroscopy. 
The details of these characterizations were reported in ref. \cite{Lee_MGBC}. 
Resistivity ($\rho$) was measured by means of a four probe method. 
Gold wires were attached to the samples with silver paste. 
In Hall resistivity measurement, gold wires were attached at the side 
edges of the crystals, in a similar way as in resistivity measurement. 
The contact resistance was less than a few Ohm. 
In order to extract the R$_H$ components, the samples were rotated 180 degree 
in a constant magnetic field. 
Raman-scattering spectra were measured in the 
pseudo-backscattering configuration with a T64000 Jobin-Yvon
triple spectrometer, equipped with a liquid-nitrogen 
cooled charge-coupled device (CCD) detector. A typical 
spectral resolution was 3 cm$^{-1}$ . The laser beam with the wavelength 
of 514.5 nm and the power of 1-6 mW was focused to a spot of
about 0.1 $\times$ 0.1 mm$^2$ on the sample surface.
To observe a SC-pair-breaking peak, 
the crystal was cleaved at room temperature before cooling. 
For the measurement of phonon-peak at room temperature, 
clean as-grown surfaces were used. 

\section{Results}

\begin{figure}
 \begin{center}
   \includegraphics[width=6cm]{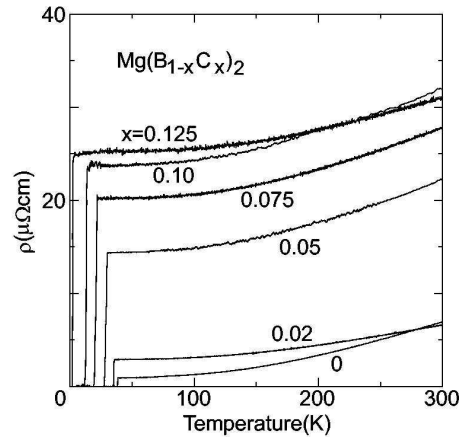}
  \end{center}
\caption{The temperature dependence of in-plane resistivity 
of Mg(B$_{1-x}$C$_x$)$_2$ single crystals. 
The T$_c$s are 38 K, 35 K, 30 K, 25 K, 10 K, and 2.0 K 
for $x$=0, 0.02, 0.05, 0.075, 0.10 
and 0.125 crystal, respectively. 
}
\label{RT}
\end{figure}

The in-plane resistivity ($\rho$) of Mg(B$_{1-x}$C$_x$)$_2$ is shown 
in Fig. \ref{RT}.
With increasing $x$, T$_c$ decreases from 38 K for $x$=0 down to 2 K 
for $x$=0.125. 
The same T$_c$-values were confirmed by the magnetization measurements 
for the same batches of crystals \cite{Lee_MGBC}. 
The most striking effect of C-substitution is the increase 
in residual  resistivity ($\rho_0$). 
The value of $\rho_0$ for $x$=0.125 is 25 times larger than that 
for $x$=0, which indicates a remarkable increase of carrier scattering 
rate due to impurity. 
Since an increase in $\rho_0$ leads to a decrease in residual resistivity 
ratio (r.r.r.), 
the result can be presented as a decrease in r.r.r. with increasing $x$. 
The empirical correlation between r.r.r. and T$_c$ in various MgB$_2$ 
samples has been reported from the very beginning of the MgB$_2$ study \cite{Buzea},  
although there is no theoretical background to consider r.r.r. 
as a parameter determining T$_c$ in an $s$-wave BCS superconductor. 
It is demonstrated in Fig. \ref{RT} that the physical meaning of r.r.r.-reduction 
is the increase in $\rho_0$, 
and that the correlation between r.r.r. and T$_c$ can be discussed 
in terms of the correlation between $\rho_0$ and T$_c$.

\begin{figure}
 \begin{center}
   \includegraphics[width=6cm]{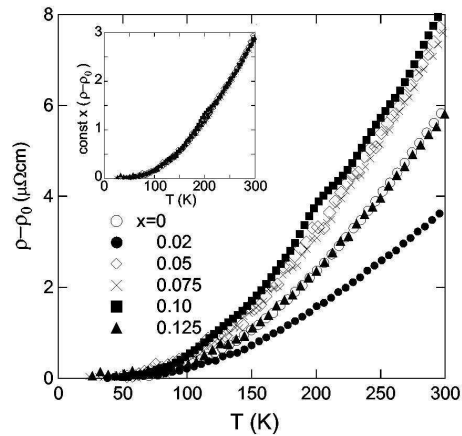}
  \end{center}
\caption{The temperature-dependent part of in-plane resistivities 
in Fig. \ref{RT}, 
obtained by subtracting the residual part. The inset shows 
the normalized T-dependent part of the resistivities. 
}
\label{R-R0}
\end{figure}

   In contrast to the dramatic increase in $\rho_0$, 
the change in resistivity slope ($d\rho/dT$) is not pronounced. 
Figure \ref{R-R0} plots the resistivity curves after subtracting residual resistivity. 
Looking at the data carefully, one may find a slight decrease in the slope 
($d\rho/dT$) at $x$=0.02. 
Similar change was observed in the MgB$_2$ single crystals with 
slightly different T$_c$ \cite{masui_ISS02}. 
By further substitution above $x$=0.02, the resistivity slope become 
steeper again. 
As it is seen in the inset of Fig. \ref{R-R0}, all the T-dependent part of 
in-plane resistivities are scaled well. 
Even in the resistivity for $x$=0.125
 T-linear temperature dependence is seen only above 200 K. 
This fact suggests that the Debye temperature estimated from resistivity 
is not a good measure of T$_c$. 

\begin{figure}
 \begin{center}
   \includegraphics[width=8.5cm]{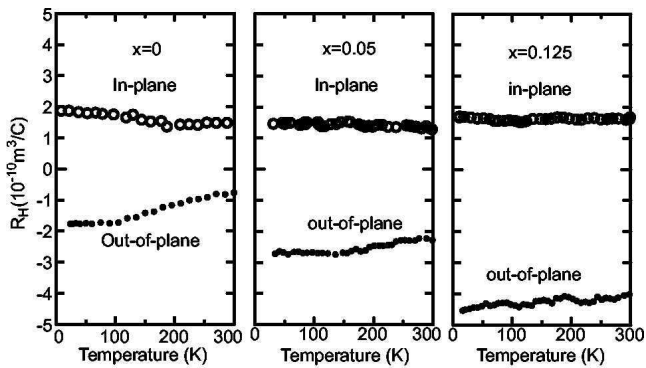}
  \end{center}
\caption{Hall coefficient for C-substituted MgB$_2$. 
In-plane R$_H$ was measured in H//c, I//ab. 
Out-of plane R$_H$ was measured in H//ab, I//ab, I$\perp$H. 
The out-of-plane R$_H$ for $x$=0 sample was replotted from 
ref. \cite{Eltsev_R_H}
}
\label{R_H}
\end{figure}

The change of carrier-concentration by C-substitution was expected 
from the valence consideration, and it should affect the R$_H$s. 
In figure \ref{R_H} the in-plane and out-of plane Hall coefficients (R$_H$) 
are presented. 
In all the samples 
the in-plane R$_H$s are positive and  the out-of-plane R$_H$s are negative. 
Although quantitative analysis of R$_H$ is difficult 
in a multi-band system, 
the negative sign of out-of-plane R$_H$ reflects the dominant contribution 
of the $\pi$-band electrons to the $c$-axis conduction, 
while the positive R$_H$ indicates the contribution of 
the $\sigma$-band holes to the in-plane conduction \cite{Eltsev_R_H}. 
The results in Fig. \ref{R_H} demonstrate that 
this two-carrier picture still holds 
even in heavily C-substituted MgB$_2$. 

As for the in-plane direction, the value of R$_H$ does not change so much, 
while the absolute value of out-of plane R$_H$ increases 
with increasing the carbon-content. 
In a simple free-carrier model, the increase of the 
absolute value of out-of-plane $R_H$ implies the decrease of electron 
carrier density, 
which is contrary to the expectation of electron doping. 
However, since the value of R$_H$ is determined by the sum of each 
contribution from $\sigma$-/$\pi$-bands, it should be carefully interpreted. 
The decrease in the $\pi$-band hole contribution might substantially effect the 
out-of-plane R$_H$ in this case. 
It should be noted that 
the calculation within a rigid-band scheme \cite{Satta} 
predicts the almost constant in-plane R$_H$ and the enhanced out-of-plane R$_H$ 
as a function of additional electron density. 
Our presented results follow the same tendency of changes 
in R$_H$s, although the absolute values of R$_H$s are one order larger than 
the calculated ones. 

\begin{figure}
 \begin{center}
   \includegraphics[width=6cm]{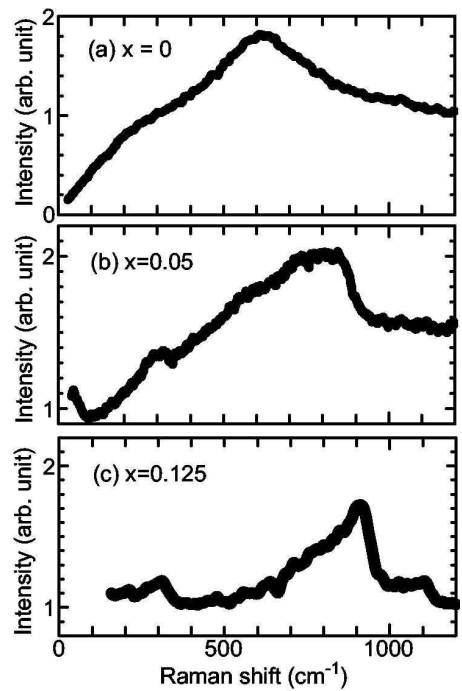}
  \end{center}
\caption{The polarized in-plane Raman scattering spectra of 
Mg(B$_{1-x}$C$_x$) at room temperature for (a) $x$=0 (replotted from ref. \cite{quilty_ISS}), (b) $x$=0.05, and (c) $x$=0.125. 
}
\label{Raman_peak}
\end{figure}

Figure \ref{Raman_peak} illustrates the Raman-scattering spectra 
for $x$=0, 0.05, and 0.125. 
With increasing $x$, the E$_{2g}$-phonon peak shows hardening. 
In the C-free MgB$_2$, the E$_{2g}$-phonon peak is extremely softened owing to 
the strong coupling with the $\sigma$-bands. 
The increase of the E$_{2g}$-peak energy suggests the weakening 
of e-ph coupling by C-substitution. 
The narrowing of the peak width is the other evidence for the weakening of e-ph coupling. 
A similar observation was reported in Al-substituted MgB$_2$ \cite{Renker,Postorino}. 
The lineshape of E$_{2g}$-peak for C-substituted MgB$_2$ becomes asymmetric 
probably because of a distribution of the 
peak energy due to the disorder induced by C-substitution. 
C-substitution also creates another peak at around 300 cm$^{-1}$ 
that may originate from the disorder. 

\begin{figure}
 \begin{center}
   \includegraphics[width=6cm]{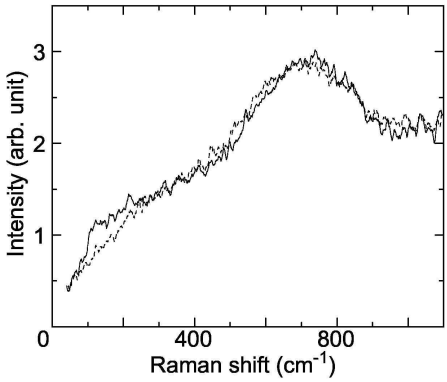}
  \end{center}
\caption{The depolarized in-plane Raman scattering spectra for $x$=0.02 
at 40 K (dotted line) and 10 K (solid line).}
\label{Raman_gap}
\end{figure}

To observe a SC gap, low temperature Raman spectra were measured. 
In figure \ref{Raman_gap}, the spectra 
for $x$=0.02 are presented. 
At T=10 K, redistribution of low-energy electronic continuum is observed,  indicating a SC-gap of 2$\Delta_\sigma$ $\sim$ 100 cm$^{-1}$. 
The sharp pair-breaking peak observed in the pure MgB$_2$ \cite{quilty_ISS}
is almost suppressed by the strong impurity scattering. 
This is consistent with the remarkable increase of residual resistivity 
shown in Fig. \ref{RT}. 
At higher C-contents, the redistribution of electronic 
continuum below T$_c$ becomes undetectable. 
The scattering rate $1/\tau$ is roughly estimated to be $\sim$90 cm$^{-1}$ 
for $x$=0.02, using $1/\tau$=30 cm$^{-1}$ for pure MgB$_2$ and 
$\rho_0$($x$=0.02)$/\rho_0$($x$=0, pure) $\sim$ 3. 
This $1/\tau$-value is comparable to 2$\Delta$($\sim$100 cm$^{-1}$). 
It means that the system changes into a dirty limit regime, 
in which a gap feature in the Raman spectra is smeared out as is typically seen in the $c$-axis Raman spectrum of MgB$_2$ \cite{quilty_c-axis}. 
The dirty-limit regime also explains the broad spectral feature for $x$=0.02. 

\begin{figure}
 \begin{center}
   \includegraphics[width=7cm]{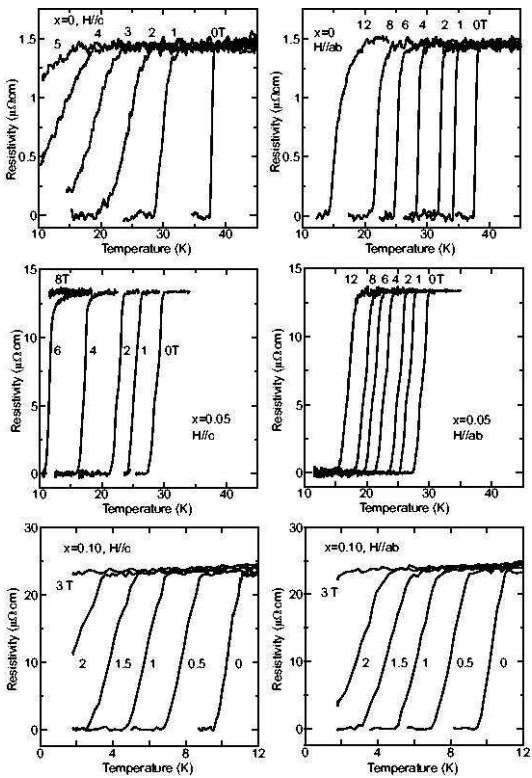}
  \end{center}
\caption{The SC transition behavior of resistivity for Mg(B$_{1-x}$C$_x$)$_2$ 
single crystals with $x$=0, 0.05, 0.10, in H//c and H//ab. 
 }
\label{transition}
\end{figure}

C-substitution also modifies superconducting state properties. 
Figure \ref{transition} shows the resistivities near T$_c$ in magnetic fields 
for $x$=0, 0.05, and 0.10.
Although the T$_c$ at H=0 decreases by C-substitution, 
the reduction of T$_c$ by magnetic field becomes small in the C-substituted crystals. 
For example, the magnetic field suppressing T$_c$ below 10 K is higher 
for $x$=0.05 than for $x$=0 in H//$c$, 
while in H//$ab$ T$_c$ at H=12 T is nearly the same for $x$=0 and 0.05 
in spite of the 10 K-difference at H=0.  
The anisotropy also decreases by C-substitution, 
and eventually disappears at $x$=0.10. 
In the crystal with $x$=0.10, the transition behavior is almost 
independent of field direction. 
The weakening of the resistivity transition broadening that is 
evident in H//$c$ also suggests the increase of three-dimensional nature 
by C-substitution. 
For $x$=0.10, SC-transitions show nearly 
parallel shifts by applying magnetic fields. 

\begin{figure}
 \begin{center}
   \includegraphics[width=6.5cm]{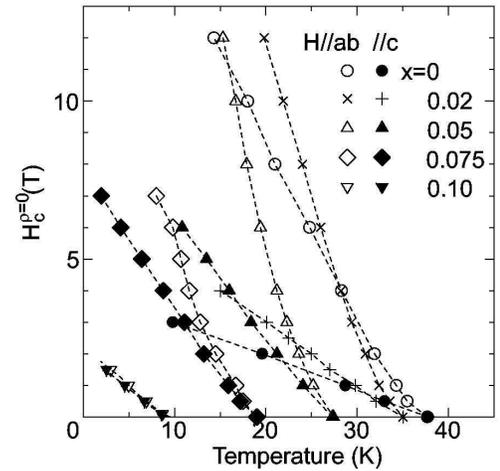}
  \end{center}
\caption{H-T phase diagram for Mg(B$_{1-x}$C$_x$)$_2$, determined 
form the zero-resistivity temperature. Lines are guides for the eyes. 
}
\label{Hc2}
\end{figure}

\begin{figure}
 \begin{center}
   \includegraphics[width=6cm]{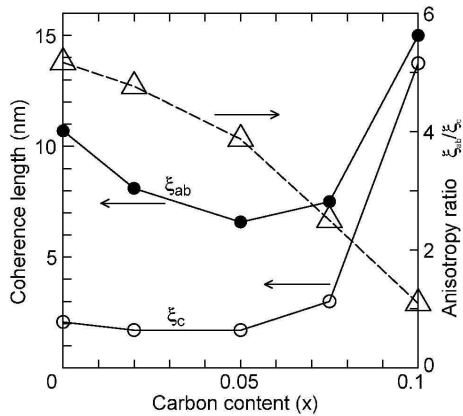}
  \end{center}
\caption{Coherence lengths are estimated by the WHH-formula, 
H$_{c2}$(0K,H//c) = $\phi_0/(2 \pi \xi^2_{ab})$ = 
0.7 ($d$H$_{c2}$/$d$T) $\times$ T$_c$ and H$_{c2}$(0K,H//ab) = $\phi_0/(2 \pi \xi_{ab}\xi_c)$ = 
0.7 ($d$H$_{c2}$/$d$T) $\times$ T$_c$\cite{WHH}. 
Here the lower-temperature slope is used 
to avoid the smaller-gap contribution. 
The open triangles represent anisotropy ratios $\xi_{ab}/\xi_c$.  
}
\label{xi}
\end{figure}

The temperature dependence of critical magnetic field H$_c^{\rho=0}$, 
determined by zero resistivity temperature, 
is plotted in Fig. \ref{Hc2}. 
Here, we regard H$_c^{\rho=0}$ as a measure of H$_{c2}$, 
although the definition of H$_{c2}$ is controversial \cite{Welp_385,Masui_mag}. 
With increasing $x$, the slope $dH_c^{\rho=0}/dT$ increases in both field 
directions. 
As a result of steep increase in the slope, 
H$_{c}^{\rho=0}$ at T=0 increases by C-substitution in both field directions, 
in spite of the T$_c$ suppression. 
This is consistent with 
the tendency in previous polycrystal studies \cite{Cheng,Soltanian,Ribeiro} 
and in our single crystal MgB$_2$ with slightly 
different T$_c$ \cite{masui_ISS02}.
In figure \ref{xi}, the coherence lengths $\xi_{ab}$ and $\xi_c$ 
estimated from the slope in Fig. \ref{Hc2} and the anisotropy ratio $\xi_{ab}/\xi_c$ are plotted. 
It is indicated that the C-substitution decreases anisotropy, 
and finally almost isotropic 
superconductivity is observed above 10 \%. 
At the low C-contents below $x$=0.075, the decrease of anisotropy ratio is 
caused by the shrink of $\xi_{ab}$ predominantly due to the shortening of 
the in-plane mean free path. 
In heavily C-substituted crystal, both $\xi_{ab}$ and $\xi_c$ are elongated. 

\section{Discussion}

The most remarkable effect of C-substitution in MgB$_2$ is the decrease 
of T$_c$. 
Since the high-temperature superconductivity of MgB$_2$ originates from the 
strong e-ph coupling in $\sigma$-bands, it is natural to ascribe the 
origin of decreasing T$_c$ to the change of $\sigma$-band properties. 
The two changes in the $\sigma$-band electronic state can be considered. 
One is the band-filling, {\it i.e.} the shift of the Fermi level,
 which decreases the hole density of states $N^\sigma$($\epsilon_F$) and 
shrinks the cylindrical Fermi surface. 
The other change is the increase of carrier scattering rate. 
It is hard to detect the change of $N^\sigma$($\epsilon_F$) selectively. 
Resistivity and Hall coefficient are affected by both of the $\sigma$- and 
$\pi$-bands, although they show anisotropy. 
In the present study, the change of $N^\sigma(\epsilon_F)$ can be seen 
in the weakening of e-ph coupling, 
because the $N(\epsilon_F)$ is one of the parameters that determines 
the coupling constant $\lambda$ \cite{an_pickett}. 
The suppression of e-ph coupling is clearly observed as the 
pronounced hardening and narrowing of E$_{2g}$-phonon in Raman scattering 
spectra in Fig.\ref{Raman_peak}. 
The change of the $\sigma$-band Fermi surface would also reduce 
the e-ph coupling via the change of deformation potential. 
We note that the shift of E$_{2g}$-phonon frequency is stronger than the 
moderate change in $N^\sigma$($\epsilon_F$) at low C-contents. 
The observed decrease of T$_c$ is not monotonous, but the 
smooth change becomes steeper as the substitution proceeds. 
This change in T$_c$-suppression rate seems to be related to 
the calculated drop of $N(\epsilon_F)$  by electron-doping \cite{an_pickett}. 
The band calculation predicts that 
superconductivity will be suppressed by $\sim$0.25 electron doping 
per unit cell \cite{an_pickett,Medvedeva}. 
The nearly suppressed T$_c\sim$2 K at $x$=0.125 in our study roughly 
corresponds to this critical doping. 

Another factor that should be discussed as an origin of decreasing T$_c$ is 
the strong interband scattering \cite{Liu} that is related to 
the large increase of residual resistivity. 
When strong scattering is introduced, 
two-gap SC picture is no longer valid, and isotropic 
superconductivity with a single SC-gap is realized. 
As for C-substituted MgB$_2$, the studies on polycrystalline samples 
have revealed that the two SC-gaps survive 
even in the sample with $T_c$ $\sim$ 20 K \cite{Ribeiro,Samuely,Schmidt}. 
It should be noted that the estimated T$_c$ for a dirty-limit is about 20 K 
without change in N($\epsilon_F$). 
Therefore, it is unlikely that the interband scattering is the primary origin of 
decreasing T$_c$ for $x\le$0.075. 
At higher C-content $x\ge$0.10, on the other hand, 
the interband scattering may play some role. 
In such highly C-substituted crystal, 
the superconductivity becomes almost isotropic, 
although the anisotropic band structure is still suggested by 
the anisotropy of Hall coefficients for $x$=0.125. 
In order to explain the isotropic superconductivity 
with the anisotropic band structure, 
the strong interband scattering scenario is preferable. 
Interband scattering may be enhanced by induced disorder by C-substitution, 
such as buckling of the B-planes. 
Accordingly, we conclude that both the decrease in $\lambda$ and 
the increase in interband scattering rate contribute to 
the decrease of T$_c$, although the degree of contribution changes 
as a function of $x$. 

In order to extract the change of the $\sigma$-bands, 
the critical field H$_{c2}$ provides useful information, because it is determined only by the $\sigma$-band gap at H$\ge$0.5 T \cite{Bouquet_single}. 
The increase of H$_{c}^{\rho=0}$ is due to the increase of 
impurity scattering via the shortening of $\xi$, 
as is expected from the relation $\xi^{-1}$ $\simeq$ $\xi_0^{-1}$ + $l^{-1}$, 
where $\xi_0$ is coherence length in clean-limit, and $l$ is mean-free path. 
The radical decrease of $l$ was clearly seen in Fig.\ref{RT}, 
as an increase of residual resistivity. 
At $x$=0.05, for example, 
the $l$ is one order shorter than that for $x$=0, 
which results in the relation $l \sim \xi$. 
The validity of dirty-limit picture was also demonstrated as 
the smearing of pair-breaking peak in the raman spectrum (Fig.\ref{Raman_gap}). 

The anisotropy change in H$_c^{\rho=0}$ can be explained 
by the anisotropic change of $\xi$. 
As seen in Fig.\ref{xi}, the anisotropy decrease is 
mainly due to the rapid shrink of $\xi_{ab}$. 
This indicates that the scattering rate for the $\sigma$-band carriers 
increases in anisotropic way. 
Since the C-substitution of atoms takes place on B-planes, it is likely 
that the introduced disorder should primarily affects 
the $\sigma$-band conduction along the B-planes. 
In the heavily C-substituted crystals, 
where the remarkable decrease in $\Delta$ elongates $\xi$, 
interband scattering may additionally contribute to the isotropic SC-state, 
as observed in the crystal with $x$=0.10 and 0.125. 

\section{Summary}
We have presented the electronic properties of 
carbon substituted MgB$_2$ single crystals. 
The effects of C-substitution are two-fold. 
One is the increase of impurity scattering, and the other is the 
band filling that reduces $N_\sigma(\epsilon_F)$ and modifies 
the shape of Fermi surface. 
The former increases rapidly the residual resistivity, enhances the the critical 
magnetic fields at T $\ll$ T$_c$, and smears out the pair-breaking peak 
in Raman spectra. 
A predominant effect of the latter is the shift of the 
E$_{2g}$-peak, suggesting the weakening of e-ph coupling by C-substitution. 
This is the primary reason for the decrease of T$_c$. 
The theoretically-expected tendency of the electron-doping effect 
on R$_H$ was confirmed, but the difference of the absolute values still 
remains to be solved. 

\section{Acknowledgements}
One of the authors (T.M) thanks to Dr. P. C. Canfield and Dr. W. E. Pickett 
for their helpful comments. 
T. M. is supported by the JSPS Research Fellowships 
for Young Scientists.
This work is supported by New Energy and Industrial Technology 
Development Organization (NEDO) as Collaborative Research and 
Development of Fundamental Technologies for Superconductivity 
Applications. 

%\newpage

%\newpage


\begin{references}
\bibitem{nagamatsu}
J. Nagamatsu, N. Nakagawa, T. Muranaka, Y. Zenitani and J. Akimitsu:
Nature \textbf{410} (2001) 63.

\bibitem{kortus}
J. Kortus, I.I. Mazin, K.D. Belashchenko, V.P. Antropov and L.L. Boyer, Phys. Rev. Lett. {\bf 86}, 4656 (2001). 

\bibitem{Uchiyama} H. Uchiyama, K.M. Shen, S. Lee, A. Damascelli, D.H. Lu, D.L. Feng, Z.-X. Shen and S. Tajima, Phys. Rev. Lett. {\bf 88}, 157002 (2002).

\bibitem{Yelland} E. A. Yelland, J. R. Cooper, A. Carrington, N. E. Hussey, P. J. Meeson, S. Lee, A. Yamamoto and S. Tajima, Phys. Rev. Lett. {\bf 88}, 217002 (2002).

\bibitem{Eltsev_R_H} Yu. Eltsev, K. Nakao, S. Lee, T. Masui, 
N. Chikumoto, S. Tajima, N. Koshizuka and M. Murakami, Phys. Rev. B 
{\bf 66} 180504(R) (2002). 

\bibitem{an_pickett} J. M. An and W. E. Pickett: Phys. Rev. Lett. 86, (2001) 4366.

\bibitem{HJ_Choi} Hyoung Joon Choi, David Roundy, Hong Sun, Marvin L. Cohen, and Steven G. Louie, Phys. Rev. B {\bf 66}, 020513(R) (2002); Nature {\bf 418} (2002), 758. 

\bibitem{Golubov} A. A. Golubov, J. Kortus, O. V. Dolgov, O. Jepsen, Y. Kong, O. K. Andersen, B. J. Gibson, K. Ahn, and R. K. Kremer, J. Phys. Condens. Matter {\bf 14}, 1353 (2002). 

\bibitem{wang}
Y. Wang, T. Plackowski and A. Junod: Physica C {\bf 355}, 179 (2001).

\bibitem{Bouquet_poly}
F. Bouquet, R. A. Fisher, N. E. Phillips, D. G. Hinks, and J. D. Jorgensen, 
Phys Rev. Lett. {\bf 87} 047001 (2001).

\bibitem{Bouquet_single} F. Bouquet, Y. Wang, I. Sheikin, T. Plackowski, and A. Junod, S. Lee, and S. Tajima, Phys. Rev. Lett. {\bf 89}, 257001 (2002).

\bibitem{Tsuda_poly}
S. Tsuda, T. Yokoya, T. Kiss, Y. Takano, K. Togano, H. Kito, H. Ihara, and S. Shin, Phys Rev. Lett. {\bf 87} 177006 (2001). 

\bibitem{quilty_c-axis}
J. W. Quilty, S. Lee, S. Tajima, and A. Yamanaka, Phys. Rev. Lett. {\bf 90} 207006 (2003). 

\bibitem{Souma} S. Souma, Y. Machida, T. Sato, T. Takahashi, H. Matsui
S.-C. Wang, H. Ding, A. Kaminski, J. C. Campuzano, S. Sasaki, and K. Kadowaki, 
Nature {\bf 423}, 65 (2003). 

\bibitem{Tsuda_single} S. Tsuda, T. Yokoya, Y. Takano, H. Kito, A. Matsushita, F. Yin, H. Harima, and S. Shin, Cond-mat/0303636 (unpublished). 

\bibitem{Simon} F. Simon, A. J\'{a}nossy, T. Feh\'{e}r, F. Mur\'{a}nyi, 
S. Garaj, L. Forr\'{o}, C. Petrovic, S. L. Bud'ko, G. Lapertot, 
V. G. Kogan, and P. C. Canfield, 
Phys. Rev. Lett. {\bf 87} 047002 (2001). 

\bibitem{Eltsev_Hc2} Yu. Eltsev, S. Lee, K. Nakao, N. Chikumoto, S. Tajima, 
N. Koshizuka and M. Murakami, Phys. Rev. B {\bf 65} 140501(R) (2001).

\bibitem{xu} M. Xu, H. Kitazawa, Y. Takano, J. Ye, K. Nishida, H. Abe, A. Matsushita and G. Kido, Appl. Phys. Lett. {\bf 79} 2779 (2001).

\bibitem{kim} 
K. H. P. Kim, J. H. Choi, C. U. Jung, P. Chowdhury, Hyun-Sook Lee, 
M. S. Park, H. J. Kim, J. Y. Kim, Z. Du, E. M. Choi, M. S. Kim, W. N. Kang,
S. I. Lee, G. Y. Sung, and J. Y. Lee, 
Phys. Rev. B. {\bf 65} 100510 (R) (2002). 

\bibitem{Pradhan_01} 
A. K. Pradhan, Z. X. Shi, M. Tokunaga,  T. Tamegai, Y. Takano, K. Togano, H. Kito and H. Ihara, Phys. Rev. B {\bf 64} 212509 (2001). 

\bibitem{sologubenko} A.V. Sologubenko, J. Jun, S. M. Kazakov, J. Karpinski, H.R. Ott, Cond-mat/0111273 (2001)

\bibitem{Slusky_Al} 
J. S. Slusky, N. Rogado, K. A. Regan, M. A. Hayward, P. Khalifah,
T. He, K. Inumaru, S. M. Loureiro, M. K. Haas, H. W. Zandbergen, 
and R. J. Cava, Nature {\bf 410}, 343 (2001).

\bibitem{Takenobu} T.Takenobu, T.Ito, D.H.Chi, K.Prassides and Y.Iwasa, Phys.Rev.B 64 (2001) 134513.

\bibitem{Ribeiro} R. A. Ribeiro, S. L. Bud'ko, C. Petrovic, and  P. C. Canfield, Physica C {\bf 384} (2003) 227.

\bibitem{Cava_Review} R.J.Cava, H.W.Zandbergen, K.Inumaru, Physica C 385 (2003) 8.

\bibitem{Liu} A. Y. Liu, I. I. Mazin, and J. Kortus, Phys. Rev. Lett. 
{\bf 87}, 087005 (2001). 

\bibitem{Pissas} M. Pissas, G. Papavassiliou, M. Karayanni, M. Fardis, I. Maurin, I. Margiolaki, K. Prassides, and C. Christides, Phys. Rev. B {\bf 65}, 184514 (2002). 

\bibitem{Samuely}P. Samuely, Z. Ho\u{l}anov\'{a}, P. Szab\'{o}, J. Ka\u{c}mar\u{c}\'{i}k, R. A. Ribeiro, S. L. Bud'ko, and P. C. Canfield, 
Phys. Rev. B {\bf 68}, 020505(R) (2003). 

\bibitem{Schmidt} H. Schmidt, K. E. Gray, D. G. Hinks, J.F. Zasadzinski, M. Avdeev, J.D. Jorgensen, and J.C. Burley, Phys. Rev. B {\bf 68}, 
060508(R) (2003). 

\bibitem{Lee_JPSJ} S. Lee, H. Mori, T. Masui, Y. Eltsev, A. Yamamoto and S. Tajima: J. Phys. Soc. Jpn.  {\bf 70}, (2001) 2255.

\bibitem{Lee_MGBC} S. Lee, T. Masui, A. Yamamoto, H. Uchiyama, and S.Tajima, 
Physica C {\bf 397}, (2003) 7. 

\bibitem{Yamamoto_pr_com} A. Yamamoto and S. Lee, private communication.

\bibitem{Buzea} C. Buzea and T. Yamashita, Superconduc. Sci. Technol. {\bf 14}, R115 (2001). 

\bibitem{masui_ISS02} T. Masui, S. Lee and S. Tajima, 
Physica C {\bf 392-396}, 281 (2003). 

\bibitem{Satta} G. Satta, G. Profeta, F. Bernardini, A. Continenza, and S. Massidda, Phys. Rev. B {\bf 64}, 104507 (2001). 

\bibitem{quilty_ISS}
J.W. Quilty, S. Lee, A. Yamamoto and S. Tajima, Physica C {\bf 378-381}, 38  (2002). 

\bibitem{Renker}B. Renker, K. B. Bohnen, R. Heid, D. Ernst, H. Schober, M. Koza, P. Adelmann, P. Schweiss, and  T. Wolf, Phys. Rev. Lett. {\bf 88}, 067001 (2002). 

\bibitem{Postorino} P. Postorino, A. Congeduti, P. Dore, A. Nucara, 
A. Bianconi, D. Di Castro, S. De Negri, and A. Saccone, 
Phys. Rev. B {\bf 65}, 020507 (2001). 

\bibitem{Welp_385} 
U. Welp, A. Rydh, G. Karapetrov, W.K. Kwok, G.W. Crabtree, C. Marcenat, 
L. M. Paulius, L. Lyard, T. Klein, J. Marcus, S. Blanchard, P.Samuely, P.Szabo, 
A.G.M. Jansen, K.H.P. Kim, C.U. Jung, H.-S. Lee, B. Kang, S.-I. Lee, 
Physica C {\bf 385}, 154 (2003). 

\bibitem{Masui_mag} T. Masui, S. Lee, and S. Tajima, Physica C {\bf 383}, 299 (2003). 

\bibitem{Cheng} C. H. Cheng, H. Zhang, Y. Zhao, Y. Feng, X. F. Rui, P. Munroe, H. M. Zeng, N. Koshizuka, M. Murakami, cond-mat/0302202.

\bibitem{Soltanian} S.Soltanian, J.Horvat, X.L.Wang, P.Munroe and S.X.Dou, Physica C {\bf 390} (2003) 185.

\bibitem{WHH} N. R. Werthamer, N. Helfand, and P. C. Hohenberg, 
Phys. Rev. {\bf 147}, 295 (1966).

\bibitem{Medvedeva} N. I. Medvedeva, A. L. Ivanovskii, J. E. Medvedeva, 
and A. J. Freeman, Phys. Rev. B {\bf 64}, 020502(R) (2001). 

\end{references}
\end{document}